# Epitaxial MgSnN$_2$ on 4H-SiC (0001): An Earth-Abundant Nitride for Green Optoelectronics and Photovoltaics

D. Gogova[1,*], D. Tran[1,2,3], V. Stanishev[1,4], D. Shafizadeh[1], C.-L. Hsiao[1], M. Kim[1,2,5], B. Pécz[6], A. Kovács[6,7], K. Frey[6], A. Sulyok[6], N. K. Singh[1,8], A. Le Febvrier[1,8], P. Eklund[1,8], and V. Darakchieva[1,2,4,5]

[1]Department of Physics, Chemistry and Biology, Linköping University, 581 83 Linköping, Sweden
[2]Center for III-Nitride Technology - C3NiT-Janzèn, Linkoping University, 581 83 Linköping, Sweden
[3]Department of Electrical Engineering, Stanford University, Stanford, CA 94305, USA
[4]NanoLund, Center for III-Nitride Technology - C3NiT-Janzèn, Terahertz Materials Analysis Center, THeMAC and Division of Solid State Physics, Lund University, 221 00 Lund, Sweden
[5]Wallenberg Initiative Materials Science for Sustainability (WISE), Department of Physics, Chemistry and Biology (IFM), Linköping University, SE-58183 Linköping, Sweden
[6]HUN-REN – Institute of Technical Physics and Materials Science, 1121 Budapest, Hungary
[7]Ernst Ruska-Centre for Microscopy and Spectroscopy with Electrons, Forschungszentrum Jülich, 52425 Jülich, Germany
[8]Department of Chemistry - Angström, Uppsala University, Lägerhyddsvägen 1, Box 538, 751 21 Uppsala, Sweden

*Corresponding author e-mail: daniela.gogova@liu.se

Group II–IV nitrides have recently emerged as a novel class of semiconductors composed of earth-abundant elements. Owing to their tunable bandgaps, comparable to those of III-nitrides, these materials are attractive candidates for replacing expensive Ga-based alloys in photovoltaics and green-gap optoelectronics. In this work, epitaxial growth of MgSnN$_2$ layers on 4H-SiC(0001) substrates by direct current magnetron sputtering is demonstrated. Mg and Sn metal targets have been co-sputtered in nitrogen-containing atmosphere at growth temperatures up to 500 °C. X-ray diffraction and cross-sectional transmission electron microscopy confirm the MgSnN$_2$ layers grow epitaxially in a wurtzite crystal structure, exhibiting the epitaxial relationships with the substrate: MgSnN$_2$ [0001]//4H-SiC [0001] and MgSnN$_2$ [10$\bar{1}$0]// 4H-SiC[10$\bar{1}$0]. Improved crystalline quality is observed for higher deposition temperatures and near-stoichiometric composition, as evidenced by the narrowing of rocking curve's linewidths. Optical characterization reveals high absorption coefficients (~$10^5$ cm$^{-1}$) in the visible spectrum, comparable to that of GaAs, highlighting the suitability of MgSnN$_2$ for photovoltaic applications. A photoluminescence emission band at ~2.4 eV is detected, highly desirable for optoelectronic devices operating in the challenging green spectral region. These results establish MgSnN$_2$ as an earth-abundant, environmentally friendly material, structurally compatible with III-nitrides, with potential for cost-efficient components in sustainable optoelectronics and photovoltaics.

**Keywords:** Group II–IV nitrides; Earth-abundant elements; epitaxial MgSnN$_2$; magnetron sputtering epitaxy; 4H-SiC (0001) substrates; optical and structural properties, green photoluminescence.

# 1. Introduction

The importance of photovoltaics and optoelectronics based on new earth-abundant materials should be seen in the perspective of the ever-growing energy demand of mankind and the need to address it with sustainable material technology. During the last decades, III-nitrides revolutionized solid state lighting and power electronics [1-6]. Expanding the nitride semiconductors family by including the II–IV ternary nitrides [7-16] offer an exciting opportunity for materials and device design that may help to overcome some of the limitations of the III-Ns, to replace the rare and very expensive gallium and indium in novel light-emitting diodes (LEDs) for the visible spectral range.

Mg(Zn)SnN$_2$ was first theoretically predicted in 2016 [17] and reported to be experimentally obtained in 2019, without any details for the method and technological parameters employed [18], demonstrating their tunability in the range of 1.45 eV to 3.5 eV. Ternary heterovalent II-IV-N$_2$, especially MgSnN$_2$, semiconductors have some advantages over group-III nitrides such as InGaN, which are the primary materials employed today in optoelectronics and photonics [1-6]: (i) composed of earth-abundant elements, *i.e.*, MgSnN$_2$ is an inexpensive material, because Mg and Sn are significantly less expensive than In and Ga and they benefit from a mature recycling infrastructure; (ii) MgSnN$_2$ has the potential to avoid phase separation in the whole alloy composition range [15], unlike InGaN [19], and (iii) MgSnN$_2$ and related II–IV–N$_2$ can be synthesized at significantly lower deposition temperatures [7-16], *i.e.*, their fabrication is more cost-effective and energy efficient than that of III-Ns. The ternary MgSnN$_2$ is also a nontoxic and stable at ambient conditions compound and hence offers additional advantages with respect to other emerging promising materials for optoelectronics and photovoltaics such as perovskites, which are either unstable (organic) or contain Pb. Furthermore, the variety of wurtzite-type semiconductors with band gaps of 1.8-2.5 eV is limited, which hinders bridging the so-called "green gap" [20] associated with the low efficiency of InGaN-based light-emitting diodes in the green spectral range. The disordered wurtzite-type MgSnN$_2$ has been shown to have a direct band gap of ~2.3 eV and it is a potential candidate for green emitters in LEDs [15].

Hypothetically, the ideal MgSnN$_2$ structure (similar to ZnSn(Si, Ge)N$_2$) can be derived from a GaN crystal by replacing every two Ga atoms with one Mg atom and one Sn atom. Thus, MgSnN$_2$ and related II−IV−N$_2$ compounds have an additional degree of freedom in comparison to III-nitrides accessible through the symmetry of the cation sublattice. The state of existing of two different valence cations (Mg$^{2+}$ and Sn$^{4+}$) offers a unique possibility to tune the material electronic, electrical and optical properties by means of controlling the degree of cation site ordering for fixed stoichiometric formula [21].

Pseudo-bulk MgSnN$_2$ in a powder form, has been synthesized by using a high-pressure-metathesis reaction [22]. MgSnN$_2$ thin films with different degrees of crystallinity and crystallographic structures (rocksalt-like, orthorhombic, and wurtzite-like) have been demonstrated by PE-MBE on yttria-stabilized zirconia (111) substrates [23] and RF-sputtering on fused silica, Si, sapphire (0002), MgO and sapphire with a GaN buffer layer (0002) substrates [14-16, 24]. Epitaxial rocksalt MgSnN$_2$ thin films were grown on MgO (111) [25]. However, the technology development is still in its infancy due to the relatively low crystalline quality of the MgSnN$_2$ thin films, motivating the need for considerable material improvement by epitaxial growth on lattice- and thermally matched substrates.

The purpose of the present paper is to demonstrate growth of high-quality epitaxial MgSnN$_2$ material and to gain insight into some basic material parameters since the epitaxial layers serve as fundamental platforms for studying the physical and chemical properties of new materials.

To achieve this aim, a substrate with crystallographic and relatively good lattice parameter matching to MgSnN$_2$ such as 4H-SiC (0001) has been selected for epitaxial material development and detailed investigations. Moreover, 4H-SiC (0001) has been used as a substrate for high-quality III-nitrides materials and devices fabrication [5, 26-28] and we are aiming at possible integration of MgSnN$_2$ in III-Ns technology.

## 2. Results and Discussion

The growth window for MgSnN$_2$ thin films with different structures (polycrystalline, textured and epitaxial) is relatively wide, i.e., the deposition temperature can be varied from room temperature to 500°C [24]. To get highly oriented material on 4H-SiC (0001) we have focused on the temperature range 350 ºC – 500 ºC. The deposition rate was determined to be 5.9-5.7 nm/min at 350°C, 5.4 nm/min at 400°C, 5.5-5.8 nm/min at 450°C and 5.3-5.6 nm/min at 500°C. The variations in the DC power ratio on the metal electrodes were used to fine tune the cation ratio (semiconductor compound composition).

### 2.1. Cation ratio

EDS measurements performed in the SEM microscope identify the cation ratio (layer composition), which varied from Sn-rich ($0.6 \leq x <1$) to stoichiometric ($x = 1$) and slightly Mg-rich ($1< x \leq 1.2$). Following growth optimization, stoichiometric MgSnN$_2$ composition corresponding to 25 at.% Mg, 25 at.% Sn and 50 at.% N was confirmed for films grown at 400°C and 450°C using an applied Mg-to-Sn target power ratio of 49 W:11 W. No other impurities were detected in the as-grown layers with the sensitivity of the EDS. However, we cannot rule out eventual oxygen contamination typical for III-nitrides grown by sputtering as well as for the other II-IV-nitride such as ZnSnN$_2$ [11]).

## 2.2. Structural properties

The films' structural properties were studied by XRD θ/2θ-scans, φ-scans, and pole-figures. A θ/2θ-scan, typical of stoichiometric $MgSnN_2$ grown at temperatures 350-450°C on 4H-SiC (0001), is illustrated in Figure 1. The intense and narrow peak at 2θ = 32.85°, can be assigned to the (0002) Bragg reflection while that at 2θ = 68.86° to the (0004) reflection of the wurtzite $MgSnN_2$ [29]. The θ/2θ scan also reveals two intense substrate peaks – the (0004) reflection of 4H-SiC is situated at 35.60°, and the (0008) one at 75.41°. In addition, three small peaks become visible between the main peaks and are attributed to SiC {000l}, which are forbidden reflections, owing to multiple scattering in the substrate. To further study the epitaxial relationship between the $MgSnN_2$ film and SiC substrate, φ-scan XRD and pole figure measurement was employed. Figure 1(b) shows azimuthal φ-scans acquired at 2θ = 44.93° and 34.07°, corresponding to the $MgSnN_2$ (10$\bar{1}$2) and 4H-SiC (10$\bar{1}$2) reflections, respectively, displayed in the upper and lower panels. During the measurements, the sample was tilted by a fixed ψ angle of approximately 44° for the film and 61° for the substrate, while the azimuthal angle φ was rotated at the selected 2θ position. Six peaks were obtained at the respective Ψ-angles with an azimuth separation of 60° in both measurements for the layer and the substrate. The φ-scans reveal excellent alignment between the $MgSnN_2$(10$\bar{1}$2) and 4H-SiC (10$\bar{1}$2). To verify if there are any other misoriented grains embedded in the film, XRD pole figures were measured at (a) 10$\bar{1}$2 $MgSnN_2$ and (b) 10$\bar{1}$2 4H-SiC reflections, displayed in Figure 1(c) and (d), respectively. Six distinct reflections located at Ψ angle ~44° and 61° can be seen in Figure 1(c) and (d), respectively, which are attributed to the six-fold symmetry of the hexagonal structure. No other distinct peaks were observed in both pole figures, indicating that the $MgSnN_2$ layer grew epitaxially on the 4H-SiC substrate. Thus, a well-defined epitaxial relationship of $MgSnN_2$ [0001]// 4H-SiC [0001] and $MgSnN_2$ [10$\bar{1}$0]// 4H-SiC [10$\bar{1}$0] can be obtained. The results confirm the epitaxial growth of c-plane–oriented $MgSnN_2$ layers with a hexagonal wurtzite structure, demonstrating structural compatibility with group-III nitrides, which is important for applications in optoelectronics. Although wurtzite $MgSnN_2$ is commonly associated with Mg-rich growth conditions due to the higher ionicity of Mg-N bonds compared to Sn-N and Zn-N bonds, we observed the wurtzite crystallographic structure even for stoichiometric and Sn-rich epitaxial layers.

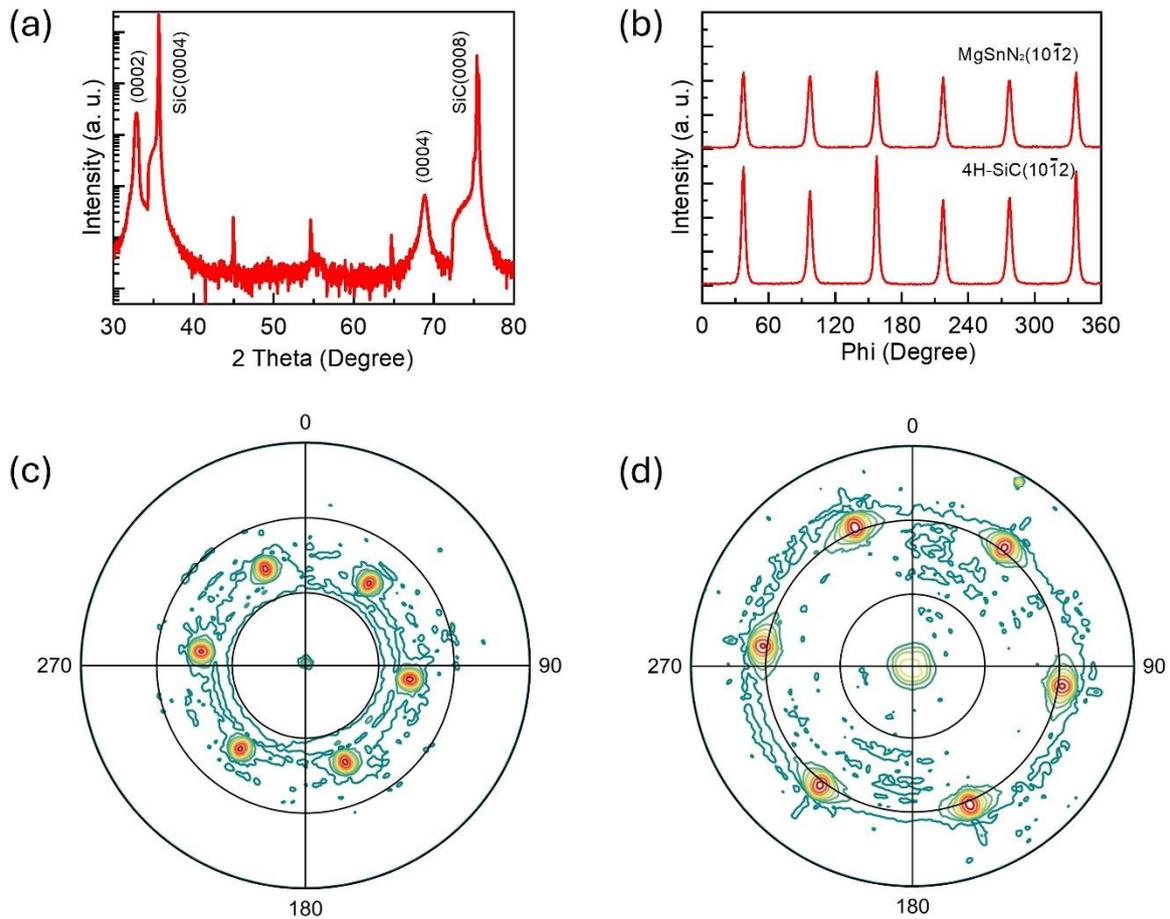

Figure 1. XRD measurements of stoichiometric $MgSnN_2$ grown on 4H-SiC (0001) substrate in the temperature range 350 °C – 450 °C. (a) θ/2θ-scan, (b) φ-scan, and pole figures of (c) $10\bar{1}2$ $MgSnN_2$ and (d) $10\bar{1}2$ 4H-SiC reflections.

The effect of the growth temperature and composition on the structural quality of the $Mg_xSn_{2-x}N_2$ epitaxial layers is illustrated in Figure 2. As obvious from Figure 2a, the layers grown in the temperature range of 350 °C - 450 °C with stoichiometric, close to stoichiometric and slightly Mg-rich composition exhibit a wurtzite (wz) phase with (0002) orientation. No other (e.g., metal, Mg-N, Sn-N) phase or additional grain orientation was observed.

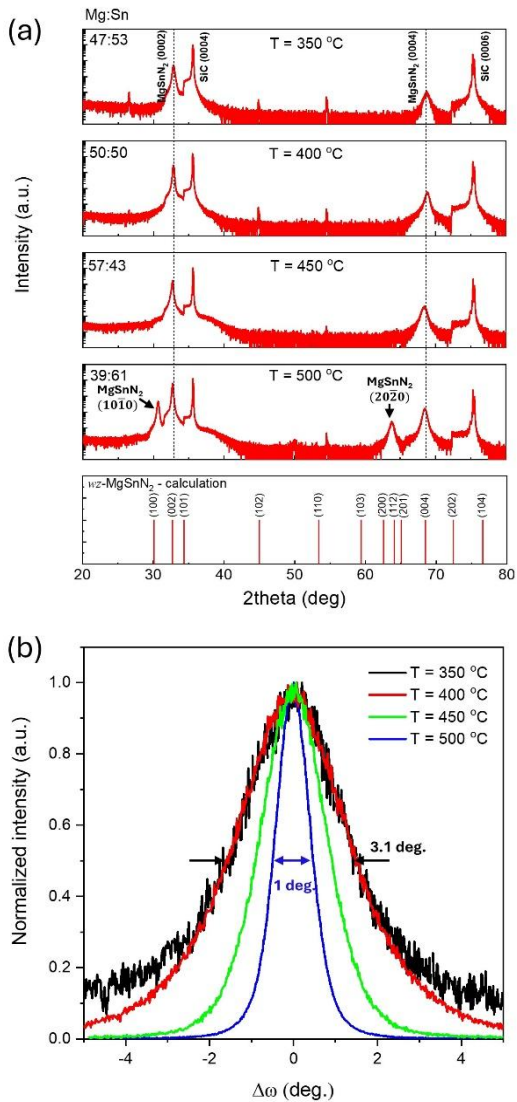

Figure 2. (a) 2θ/θ scans of the $wz$-Mg$_x$Sn$_{2-x}$N$_2$ layers grown at temperatures in the range of 350 ºC - 500 ºC, the XRD spectrum theoretically calculated for stoichiometric MgSnN$_2$ is also shown, (b) $(10\bar{1}2)$ RCs of the corresponding layers reported in (a).

At a growth temperature of 500 ºC and Sn-rich layer composition, the main growth direction remains unchanged; however, the layer incorporates additional m-plane oriented crystal domains, indicated by the presence of ($10\bar{1}0$) and ($20\bar{2}0$) diffraction peaks at 2θ of 30.6º and 63.7º, respectively. The lattice constants of the c-plane crystals, measured using the (0002) and ($10\bar{1}2$) diffractions, are found to be $a$ = 0.349 and $c$ = 0.547 nm. Calculations using both ($10\bar{1}0$) and ($20\bar{2}0$) diffractions yield the same value of $a$ = 0.337 nm for the m-plane crystals. At this high temperature, the layer becomes more Sn-rich with Mg:Sn cation ratio of 39:61. This behavior is attributed to the increased volatility of Mg at higher growth temperatures.

Despite the large variation in the layers' cation ratio with growth temperature, the XRD peaks of (0002) and (0004) diffractions remain almost unchanged and aligned well with the corresponding peak positions of the powder diffraction pattern for stoichiometric $MgSnN_2$. The XRD powder diffraction pattern was generated by us employing Vesta (3D visualization program designed for crystallography, materials science, and chemistry), with cell parameters of the disordered wz-$MgSnN_2$ obtained from electronic calculation and structural optimization using density functional theory, yielding $a$ = 0.3426 and $c$ = 0.5474 nm [29]. Notably, the ($10\bar{1}0$) and ($20\bar{2}0$) peak positions significantly differ from those of the reference pattern by +0.52º and +1.3º, due to the differences in the lattice constant $a$. Interestingly, the lattice constant $a$ of the c-plane and m-plane domains in the layer are significantly different, which could be associated with differences in (i) the strain states and/or (ii) the cation ratio of these two different types of crystal domains. Figure 2b presents the ($10\bar{1}2$) rocking curves (RCs) of the $Mg_xSn_{2-x}N_2$ layers. The RC full width at half maximum (FWHM) is as large as 3.1º for the layer grown at T = 350 ºC, while higher growth temperatures result in progressively narrower linewidths. In particular, the minimum FWHM of 1º is obtained for the layer grown at T = 500 ºC. The result indicates better crystalline quality of the individual domains at higher growth temperatures up to 500 ºC. Due to the appearance of m-plane crystal phase orientation for the Sn-rich sample and the enhanced Mg volatility at higher temperatures the deposition temperature was not further increased in our series of experiments. No RCs for $Mg_xSn_{2-x}N_2$ epitaxial layers have been reported in literature so far to have a basis for comparison.

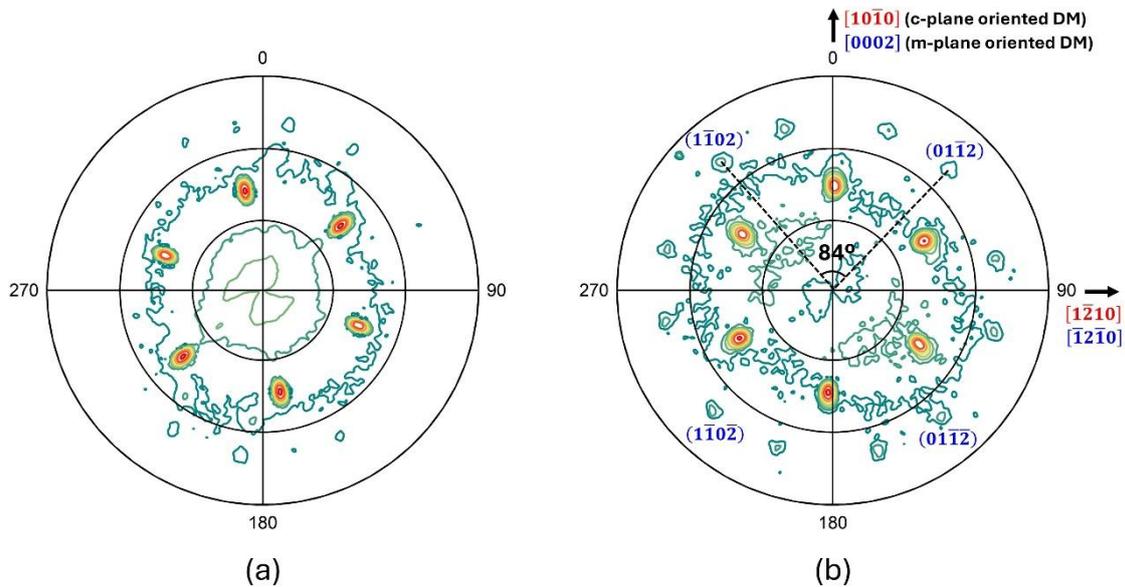

Figure 3 Pole figures of $\{10\bar{1}2\}$ diffractions measured for the MgSnN$_2$ layers grown at two different temperatures of T = 450 ºC and cation ratio 57:43 - (a) and T = 500 ºC, cation ratio of 39:61 - (b). The mutual orientations marked for the c-plane and m-plane oriented domains are in red and blue, respectively.

The pole figure of the layer grown at T = 450 ºC and cation ratio of 57:43 contains six peaks at an inclination angle of 44º, corresponding to $\{10\bar{1}2\}$ reflections of c-plane-oriented wurtzite crystal, revealing the expected six-fold symmetry of the lattice (see Figure 3a). The layer grown at T = 500 ºC and cation ratio of 39:61 additionally shows 12 peaks at an inclination angle of $\Psi = 71.5º$, which have significantly lower intensity compared to the main peaks. The extra peaks match the $\{10\bar{1}2\}$ reflections characteristic of m-plane (1-100) oriented crystals, indicating the coexistence of both c-plane and m-plane domains in the layer grown at 500 ºC, with the m-plane component being minor (Figure 3b). For the m-plane domains, the measured reflections particularly correspond to (01$\bar{1}$2), (1$\bar{1}$02), (01$\bar{1}\bar{2}$) and (1$\bar{1}$0$\bar{2}$) planes. The first and second reflections form an azimuth angle of 84º, while it is 96º for the first and third ones. The full set of 12 peaks arises from a 60º rotation repetition of the m-plane oriented domain. Notably, the results indicate that the c-axis of the m-plane domains aligns along the $\langle 10\bar{1}0 \rangle$ direction of the c-plane host domains.

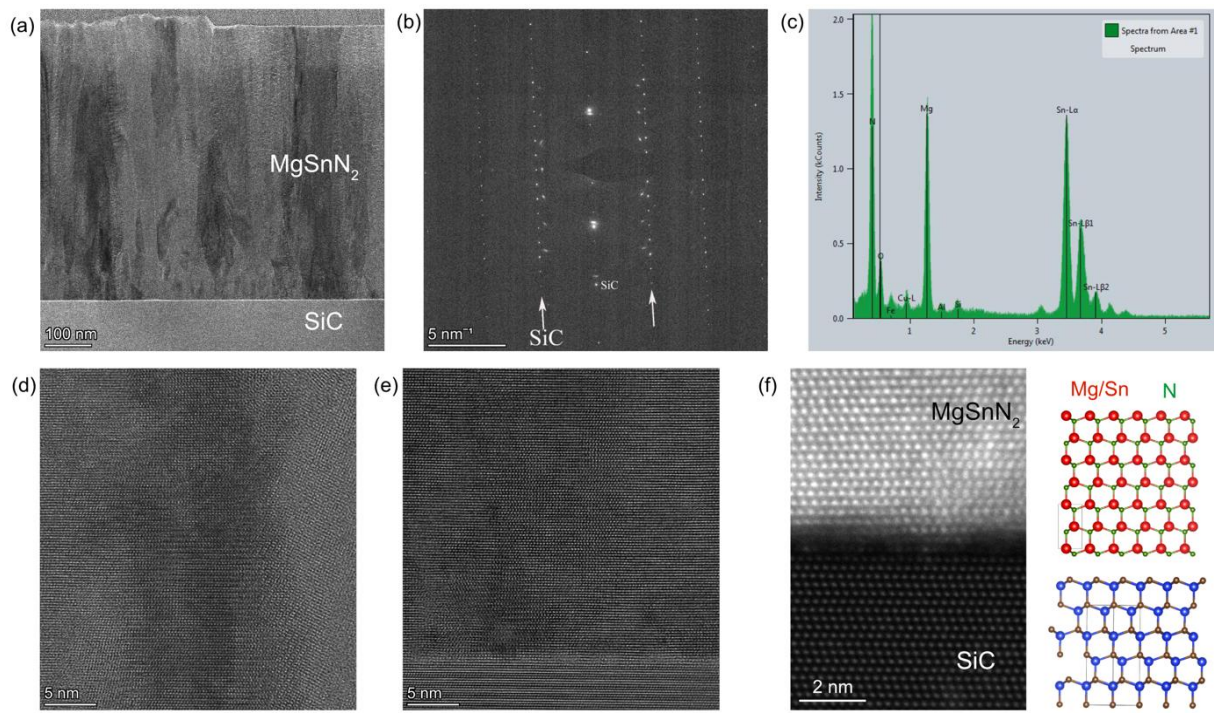

Fig. 4 (a) BF STEM image showing the columnar structure of the $MgSnN_2$ layer on 4H-SiC substrate. The Pt/C protective layer deposited during the FIB preparation is visible on the top. (b) SAED pattern recorded from the interface region showing the epitaxial relationship of wz-$MgSnN_2$ and 4H-SiC (0001). (c) EDS spectrum recorded from the $MgSnN_2$ layer. (d,e) BF STEM images of the $MgSnN_2$ layer and interface regions, respectively. (f) HAADF STEM image of the interface and the atomic arrangements of wz-$MgSnN_2$ and 4H-SiC.

The structure of the stoichiometric $MgSnN_2$ layer grown at temperature of 400 ºC was further studied in cross-sectional geometry as illustrated in Figure 4, by preparing a thin specimen using FIB. The BF STEM image in Figure 4(a) shows the columnar structure of the MgSnN$_2$ layer, which nucleates at the 4H-SiC (0001) surface. The layer thickness is approximately 600 nm, and the lateral dimension of the columns is in the range of 60 nm to 80 nm. The epitaxial relationship between MgSnN$_2$ and 4H-SiC is confirmed as MgSnN$_2$[0001]//4H-SiC [0001] and MgSnN$_2$ [10$\bar{1}$0]// 4H-SiC [10$\bar{1}$0] using a selected area electron diffraction (SAED) pattern recorded from the interface region. Chemical composition measurement using EDS (Figure 4(c)) reveals a composition of Mg - 28 at. %, Sn - 25 at. % and N - 47 at. %. These values are in good agreement with the EDS measurement obtained from the bulk sample using SEM. Figure 4(d) shows a high-resolution BF STEM image of the MgSnN$_2$ layer acquired from the mid-section confirming the single crystalline nature of the column. The interface structure is presented in Figure 4(e), where MgSnN$_2$ grains are observed to nucleate and grow epitaxially on the 4H-SiC substrate. The interface between the substrate and the layer was further investigated with element sensitive high-angle annular dark-field (HAADF) STEM imaging, as shown in Figure 4(f), where the intensity scales approximately as I~ $Z^2$. A 1-2 ML thick region of dark contrast at the interface indicates a high concentration of light elements such as C, O or N. Next to the HAADF STEM image, the unit cells of the wz-MgSnN$_2$ and 4H-SiC (0001) are shown to illustrate the epitaxial relationship.

## 2.2 Optical properties

Figure 5 shows the absorption coefficient spectra, extracted from spectroscopic ellipsometry measurements, for a set of wurtzite highly oriented MgSnN$_2$ layers with different cation ratios. The absorption coefficients for all layers are in the high $10^4$ range up to ~ $10^5$ cm$^{-1}$ in the visible spectrum, i.e., values approaching these of the solar-energy-relevant GaAs [9]. Recent theoretically calculated absorption coefficient and reflectivity curves of MgSnN$_2$ validate that the disordered rock-salt structure might be used as an absorber layer in the higher energy region of the visible range of spectrum in tandem solar devices [9], whereas the disordered wurtzite and orthorhombic crystal structures could serve as window layers of solar cells. However, our results demonstrate that highly ordered epitaxial MgSnN$_2$ layers with a wurtzite structure have large absorption coefficients too.

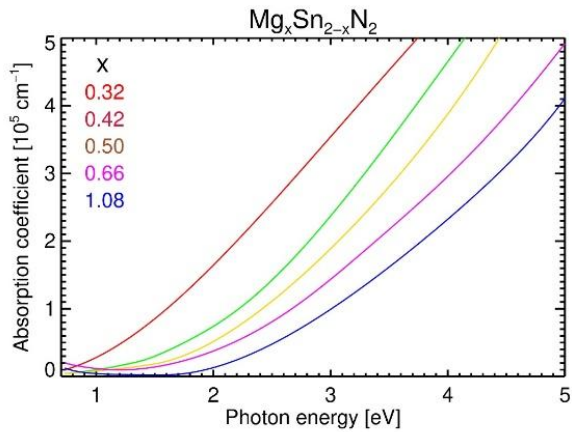

Fig. 5 Absorption coefficient spectra vs photon energy of a set of wurtzite highly oriented Mg$_x$Sn$_{2-x}$N$_2$ layers with different stoichiometry (x=0.32; 0.42; 0,50; 0.66; 1.08).

The bandgap determination of the stoichiometric wz-MgSnN$_2$ was conducted by reflection electron energy loss spectroscopy (REELS) technique. Prior to the measurements, the oxygen adsorbed on the surface of the as-grown samples, kept at ambient conditions, was removed in several steps via Ar ion sputtering controlled by Auger electron spectroscopy. A typical REELS spectrum of a stoichiometric MgSnN$_2$ layer deposited onto 4H-SiC (0001) substrate is depicted in Figure 6. Since the elastic peak is three orders of magnitude more intense, accurately measuring the low-intensity loss region of the spectrum required extended averaging. The average of ten spectra was used for the MgSnN$_2$ band gap estimation.

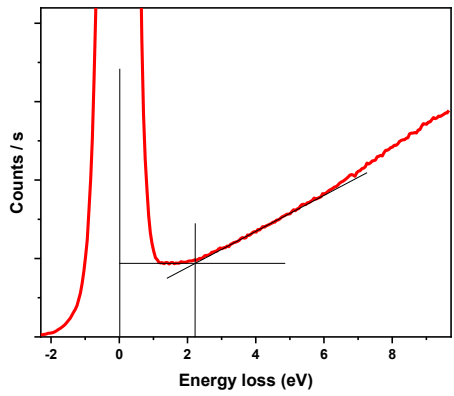

Fig. 6 A typical REELS spectrum of a stoichiometric MgSnN$_2$ layer deposited on 4H-SiC (0001) substrate.

The band gap value of epitaxial material was determined to be 2.24 ±0.1 eV. Optical band gap values ranging from 2.27 to 2.43 eV have been reported in the literature for the direct bandgap stoichiometric MgSnN$_2$ thin films with a different degree of crystallinity [13-15, 24].

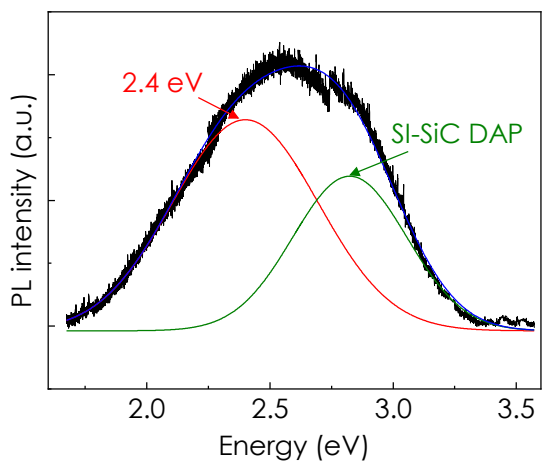

Fig. 7 A Low-temperature photoluminescence spectrum typical of epitaxial MgSnN$_2$ sample.

A low-temperature photoluminescence spectrum, measured at 80 K with an UV laser excitation of 351 nm, typical of a MgSnN$_2$ sample is illustrated in Figure 7. The PL peak at 2.85 eV is identified as the N-Al donor-acceptor pair (DAP) emission of SI-SiC [33] appearing at low-temperatures. The PL emission band with a maximum at 2.4 eV is due to the MgSnN$_2$. Obviously, the material grown has a green-light photoluminescence with a peak maximum at 2.4 eV and therefore is promising for a source in optically active devices for the critical up-to-date green gap [20]. Direct bandgap semiconductors with bandgaps in this range are strongly needed for the development of green LEDs since the large Indium (In) content in the present green light InGaN-based LEDs leads to decreased efficiency due to the In-phase segregation appearing at large In contents in the InGaN alloy [19]. The indium segregation principally originates from the large lattice mismatch between InN and GaN.

Additionally, the wurtzite MgSnN$_2$ (0001) could be integrated with the well-developed III-nitride semiconductors because of their structural compatibility. Moreover, such direct bandgaps [15, 29] are required for top cells in high-efficiency tandem solar cells.

Electrical Hall effect measurements performed on a series of MgSnN$_2$ samples showed high electron concentration reaching $4.7 \times 10^{19}$ cm$^{-3}$ for stoichiometric films and $4.3 \times 10^{20}$ cm$^{-3}$ for Sn-rich layers. The corresponding electron mobilities range from 3.3 to 5.9 cm²/V·s. These values are consistent with earlier reports in literature [22, 29, 30].

First principle calculations have shown that wurtzite MgSnN$_2$ exhibits self-doped *n*-type conductivity, and the antisite defect Sn$_{Mg}$ was suggested to be the primary source of free electrons [31]. The propensity for forming the stoichiometry preserving Mg$_{Sn}$+ Sn$_{Mg}$ defect complex leads to cation disorder in MgSnN$_2$, which induces a band-gap reduction because of a violation of the octet rule [32]. The cation-size mismatch has been suggested to be the reason for structural distortion, configurational disorder and valence-band splitting in II-IV-N$_2$ semiconductors [32]. However, the unintentional high free electron concentration could be also due to oxygen contamination, which is inevitably present in nitride semiconductors and has been detected in sputtered ZnSnN$_2$ by SIMS [11].

The unintentionally high electron concentration in as-grown MgSnN$_2$ still remains a scientific and technological challenge and additional study is needed to clarify its nature.

3. **Conclusions**

We have demonstrated magnetron sputter epitaxy of wurtzite MgSnN$_2$ on 4H-SiC (0001) substrates and have proven its structural compatibility with group-III nitrides. X-ray diffraction and cross-sectional TEM measurements verified the layers are epitaxial of hexagonal phase, exhibiting epitaxial relationship with the substrate described by: MgSnN$_2$[0001]//4H-SiC [0001] and MgSnN$_2$ [10$\bar{1}$0]// 4H-SiC[10$\bar{1}$0]. Crystalline quality improves by increasing the deposition temperature up to ≈ 500ºC and for stoichiometric composition. The calculated absorption coefficient values reach ≈10$^5$ cm$^{-1}$ in the visible range of spectrum, i.e., relevant to photovoltaic applications as an absorber in tandem solar cells. The material grown manifests a green-light photoluminescence with a peak maximum at ≈2.4 eV, which indicates it could be advantageous for a source in optically active devices in the critical up-to-date green gap of the spectrum.

4. **Experimental section**

Mg$_x$Sn$_{2-x}$N$_2$ thin films with a wide compositional x-range (0.6 ≤ x ≤ 1.2) were deposited by a DC reactive magnetron co-sputtering of metal Mg (99.995 vol.%) and Sn (99.995 vol.%) targets in N$_2$-containing atmosphere employing an ultra-high vacuum Magnetron Sputter deposition chamber with a base pressure of ∼ 5.33x10$^{-7}$ Pa. (see for details [34]). The magnetron assemblies were situated on the bottom lid of the chamber each 90° degrees in a co-focal position with an off-axis position of 30°. The target diameter was 50 mm and the distance between the targets and the substrate holder was approximately 135 mm [34]. For all deposition experiments a mixture of $N_2$ (99.999%) and Ar (99.999%) was used as a sputter gas in the thin film synthesis. The substrates with lateral sizes 10x10 mm or 10x5 mm were cleaned in an ultrasonic bath as follows: 5 min in acetone, 5 min in propanol, 5 min in deionized water and finally dried in a $N_2$-flow of high purity. $MgSnN_2$ films were co-sputtered on 4H-SiC (0001). The deposition temperature in a series of experiments varied from 200 to 500°C with a step of 50°C, in nitrogen-containing atmosphere with Ar addition. The $N_2$ flow was 72 sccm, the Ar flow 18 sccm and the sputtering pressure was fixed at 0.48 Pa. The $MgSnN_2$ films prepared were nominally undoped. The crystalline structures and epitaxial relationship were analyzed by θ/2θ-scan, φ-scan, and pole-figure x-ray diffraction (XRD) employing a PANalytical Empyrean X-ray diffractometer with a Cu-Kα source and a parabolic mirror as the primary optics and a 0.27° parallel plate collimator as the secondary optics. The thin film thickness was determined by means of a Scanning electron Microscope (SEM) Zeiss, Sigma 300, while the films composition – by Energy dispersive X-ray spectroscopy (EDS) in the SEM.

The cross-sectional specimens for transmission electron microscopy (TEM) studies were prepared using focused Ga ion beam (FIB) sputtering in a dual beam scanning electron microscope (ThermoFisher Scios 2) following a conventional lift-out method. To study the structure and epitaxy of $MgSnN_2$ films, electron diffraction patterns and bright field images were recorded using an aberration-corrected electron microscope (ThermoFisher Themis 200) operated at 200 kV acceleration voltage. The chemical element distribution was measured by EDS and scanning TEM (STEM). The TEM data was processed using Velox software (ThermoFisher).

Reflection electron energy loss spectroscopy (REELS) for band gap measurement was carried on layers grown on 4H-SiC. Since the REELS measurement is a surface sensitive technique an oxidized region at the sample surface was removed by ion sputtering before the REELS spectra were recorded. For this purpose, low energy ions sputtering was carried out with 1keV Ar ion beam scanned over 2 mm spot at the surface. The REELS spectra were detected using an Escalab Xi+ (Thermo Fisher) equipment and its low temperature electron source with low energy width of the exciting beam was used. Measurement took place at 1 keV primary energy detecting 300 μm spot of the surface observing. Both primary electrons and escaping electrons are related to the surface plane at normal angle conditions. The spectra were detected at 0.05 eV steps with 0.5 eV energy resolution of the analyzer.

Spectroscopic ellipsometry (SE) using a dual-rotating compensator ellipsometer (RC2, J.A. Woollam Co., Inc.) was performed in the range from 0.75 to 5.9 eV for the characterization of sample optical properties]. Psi and Delta spectra were measured at angles of incidence from 45° to 75° with a step of 5°. The experimental data are analysed using the WVase32 software (J.A. Woollam Co. Inc.) employing a stratified optical model with parametrized model dielectric functions (MDFs) assigned to each layer. The MDF of 4H-SiC was determined from measurements of a bare substrate and was kept fixed during the analysis. The MDF of the $MgSnN_2$ film was modeled as a sum of one Tauc-Lorentz and two Lorentz oscillators. The film thickness and the parameters of the oscillators were determined by non-linear least-squares fit to the experimental data. The absorption coefficients of the $MgSnN_2$ films were then obtained from the respective best-matched MDFs. Low-temperature photoluminescence measurements were carried out at 80 K by cooling samples in a liquid-nitrogen cryostat. The layers were excited using the 351 nm ultraviolet line of an Ar-ion laser with a power below 5 mW, moderately focused onto the sample to a spot diameter of approximately 100 μm. The emitted PL signal was analyzed using a monochromator (Jobin Yvon HR460) equipped with a multichannel detector.

**Acknowledgements**

We thank L. Illés (HUN-REN) for FIB preparation. This work is supported by the Wallenberg Initiative Materials Science for Sustainability (WISE) funded by the Knut and Alice Wallenberg Foundation, the Swedish Government Strategic Research Area in Materials Science on Functional Materials at Linköping University (Faculty Grant SFO-Mat-LiU No. 2009 00971), the Knut and Alice Wallenberg foundation through the Wallenberg Academy Fellows program (KAW-2020.0196, P.E.), the Swedish Research Council (VR) under Project Nos. 2021-03826 (P.E.), 2025-03680 (P. E.), and 2025-03760 (A. F.). This work is supported by the Swedish Research Council (VR) under Grant No. 2023-04993, and by the Swedish Government Strategic Research Area NanoLund and in Materials Science on Functional Materials at Linköping University, Faculty Grant SFO Mat LiU No. 009-00971. V.D. acknowledges support by the Knut and Alice Wallenberg Foundation for a Scholar award (Grant No. 2023.0349), and D.Q.T acknowledges support by the Knut and Alice Wallenberg Foundation for a Postdoctoral Fellowship to Stanford University. B. Pecz acknowledges the support of the NKFIH Advanced 152987 project and A.K. acknowledges the support of the MTA Distinguished Fellowship Program 2024. D.G. thanks the Carl Trygger Foundation for Scientific Research (grant No. 312449).

**Data Availability Statement**

The data that supports the findings of this study are available from the corresponding author, [D.G.], upon reasonable request.

**Conflict of Interest**

All authors declare no conflicts of financial/commercial interests.